\documentclass{article}
\usepackage[a4paper, total={6in, 8in}]{geometry}
\usepackage{amssymb}
\usepackage{graphicx,epstopdf}
\usepackage{gensymb}
\usepackage{epsfig}
\usepackage{color}
\usepackage{epstopdf}
\usepackage{amsmath}
\usepackage{caption}
\usepackage{subcaption}
\usepackage{authblk}
\usepackage[super,comma]{natbib}
\usepackage{url}

\newcommand{\grl}{    {Geophys. Res. Lett.}}
\newcommand{\jgr}{    {J. Geophys. Res.}}
\newcommand{\ssr}{    {Space Sci. Rev.}}

\newcommand{\apj}{ {Astrophys. J. }}

\newcommand{\pre}{ {Phys. Rev. E}}
\newcommand{\prl}{ {Phys. Rev. Lett.}}
\newcommand{\apjl}{ {Astrophys. J. Lett. }}

\newcommand{\area}{\mathcal{S}}
\newcommand{\h}{\mathcal{H}}
\newcommand{\p}{\tilde{p}}
\newcommand{\s}{\tilde{s}}
\def\mP{{\cal P}}

\begin{document}


\title{Electron resonant interaction with whistler-mode waves around the Earth's bow shock II: the mapping technique} 

\author[1,3]{David S. Tonoian}
\author[2]{Xiaofei Shi}
\author[2,4]{Anton V. Artemyev}
\author[1,2]{Xiao-Jia Zhang}
\author[2]{Vassilis Angelopoulos}
\affil[1]{Department of Physics, University of Texas at Dallas, Richardson, TX, USA; david.tonoian@utdallas.edu}
\affil[2]{Department of Earth, Planetary, and Space Sciences, University of California, Los Angeles, USA; sxf1698@g.ucla.edu}
\affil[3]{Faculty of Physics, National Research University Higher School of Economics, Moscow, Russia, 105066.}
\affil[4]{Space Research Institute, RAS, Moscow, Russia}

\maketitle

\begin{abstract}
Electron resonant scattering by high-frequency electromagnetic whistler-mode waves has been proposed as a mechanism for solar wind electron scattering and pre-acceleration to energies that enable them to participate in shock drift acceleration around the Earth's bow shock. However, observed whistler-mode waves are often sufficiently intense to resonate with electrons nonlinearly, which prohibits the application of quasi-linear diffusion theory. This is the second of two accompanying papers devoted to developing a new theoretical approach for quantifying the electron distribution evolution subject to multiple resonant interactions with intense whistler-mode wave-packets. In the first paper, we described a probabilistic approach, applicable to systems with short wave-packets. For such systems, nonlinear resonant effects can be treated by diffusion theory, but with diffusion rates different from those of quasi-linear diffusion. In this paper we generalize this approach by merging it with a mapping technique. This technique can be used to model the electron distribution evolution in the presence of significantly non-diffusive resonant scattering by intense long wave-packets. We verify our technique by comparing its predictions with results from a numerical integration approach.
\end{abstract}

\maketitle

\section{Introduction}
The high-frequency electromagnetic whistler-mode wave is one of the most intense wave modes that can resonate with solar wind electrons and provide electron pitch-angle and energy scattering \cite{Tong19:ApJ,Cattell&Vo21,Cattell21:scattering,Vo22}. In particular, intense whistler-mode waves are observed around interplanetary shocks \cite{Wilson13:waves,Davis&Cattell21} and Earth's bow shock \cite{Hull12,Hull20,Page&Vasko21,Shi22:ApJ}. Resonant interactions between electrons and such intense waves may result in strong pitch-angle scattering \cite{Oka17,Oka19,Shi20:foreshock_whistlers}, which is important for electron participation in the stochastic-shock-drift acceleration \cite{Amano20,Amano22:review}. Moreover, in the presence of ambient magnetic field gradients and shock-related electrostatic fields, whistler-mode waves may directly accelerate solar wind electrons \cite{Kuramitsu05,Artemyev22:jgr:bowshock}. This important role of whistler-mode waves for suprathermal electron dynamics drives many theoretical investigations of wave-particle resonant interactions around shocks (see discussions in Ref. \cite{Wilson14,Amano20}).

The basic theoretical approach for modeling wave-particle resonant interactions, the quasi-linear theory \cite{Vedenov62,Drummond&Pines62}, can be used to estimate the electron scattering rates provided by whistler-mode waves \cite{Veltri&Zimbardo93,Amano20}. However, the criteria for this theory (low wave intensity, broad wave spectrum; see Ref. \cite{Karpman74:ssr,Shapiro&Sagdeev97}) can be violated around the shock, where the observed whistler-mode waves are very intense and sufficiently narrow-band \cite{Hull20,Shi22:ApJ,Artemyev22:jgr:bowshock}. Such waves may resonate with electrons nonlinearly. Presently, models of nonlinear resonant interactions are mostly developed for the Earth's inner magnetosphere (see, e.g., Refs. \cite{Shklyar09:review,Albert13:AGU,Hsieh&Omura17,Artemyev21:jpp} and references therein) and require significant modifications to be suitable for the plasma and wave characteristics commonly observed around the Earth's bow shock. In this paper we seek to advance such models for planetary shocks, and in particular Earth's bow shock, their most accessible representative.

\begin{figure*}
\centering
  \begin{subfigure}{0.47\textwidth}
    \includegraphics[width=1\textwidth]{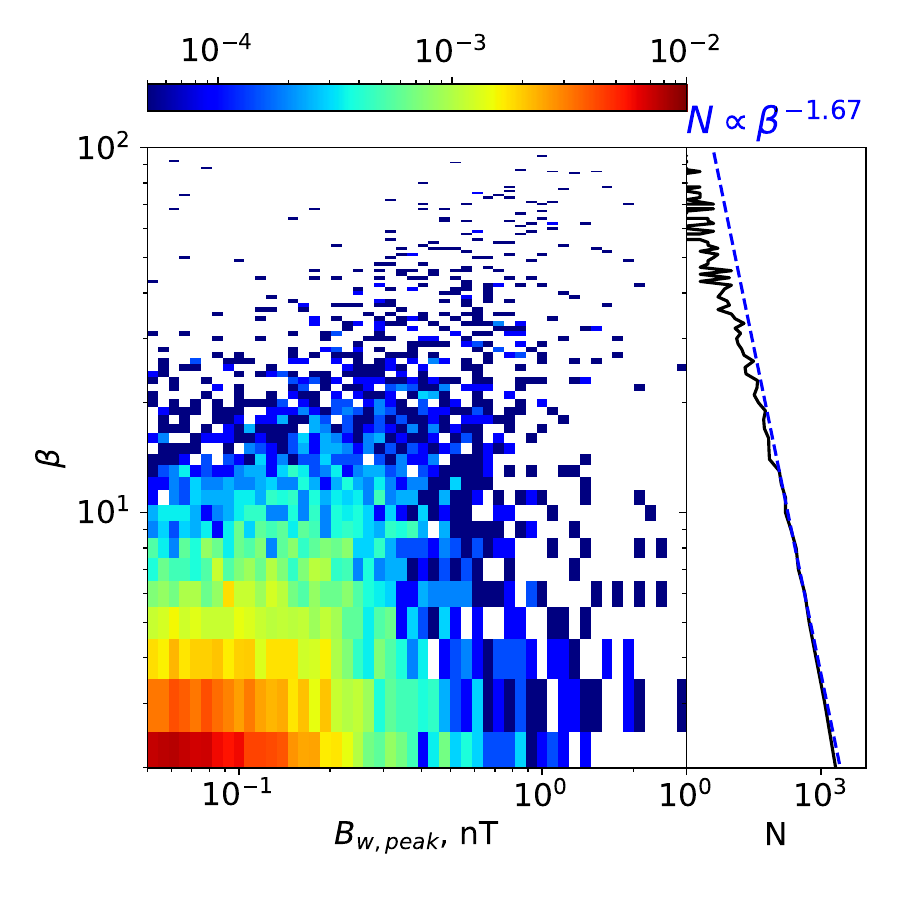}
    \end{subfigure}
  \begin{subfigure}{0.47\textwidth}
    \includegraphics[width=1\textwidth]{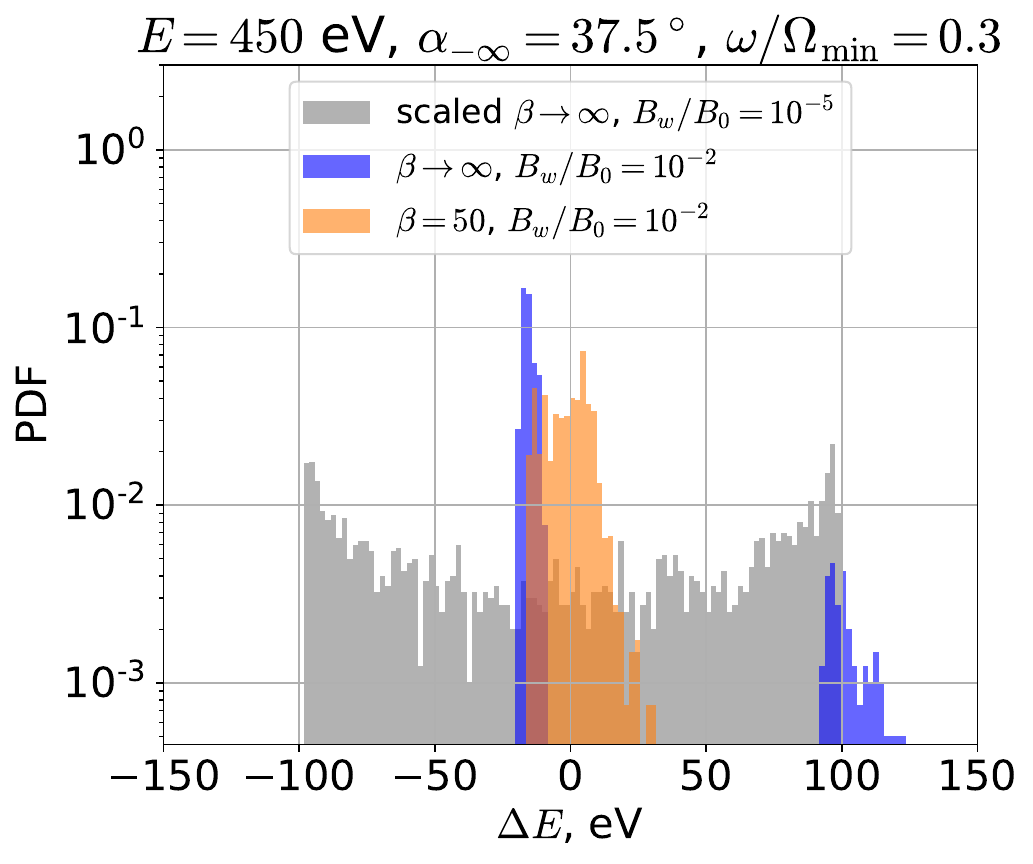}
    \end{subfigure}
\caption{(left) The distribution of waves in $(B_w,\beta)$ space for the foreshock region (see details of wave statistics in Ref. \cite{Shi22:ApJ}). (right)
Three examples of probability distributions of energy change $\Delta E$ for a single resonant interaction of electrons (having the same initial energy/pitch-angle): (grey) low-intensity wave-packet, (blue) long intense wave-packet, and (orange) short intense wave-packet. The wave magnetic field is normalized to the minimum value of the ambient magnetic field (see details in the text), and for the distribution with $B_w/B_0=10^{-5}$, we rescale $\Delta E \to \Delta E\cdot 10^{3}$ to compare with the other two $\Delta E$-distributions evaluated for $B_w/B_0=10^{-2}$. 
\label{fig1} } 
\end{figure*}

Intense whistler-mode waves detected at the bow shock and in the foreshock region are strongly modulated by low-frequency compressional waves \cite{Hull12,Hull20}, i.e., whistler-mode waves are propagating in short wave-packets. Examples of such wave-packets are shown in Figure \ref{fig1} of Ref. {\it Electron resonant interaction with whistler-mode waves around the Earth's bow shock I: the probabilistic approach}, hereinafter referred to as Paper 1. Figure \ref{fig1}(left panel) shows the probability distribution of whistler-mode waves in the space of wave amplitude $B_w$, in $nT$, and wave-packet size $\beta$, in wave periods. Although there are long and intense wave-packets with $B_w>100$pT and $\beta\geq 20$, the majority of the observed intense waves propagate in short wave-packets, characterized by $\beta<20$. Such a strong wave modulation  (short packet size) can significantly reduce the efficiency of nonlinear resonant interactions\cite{Tao13,Allanson20,Allanson21}, because electrons spend less time in resonant interactions with short wave-packets and cannot gain substantial energy (see simulation results in Refs. \cite{Mourenas18:jgr,Zhang20:grl:phase,An22:Tao,Gan22}). 

Figure \ref{fig1}(right panel) illustrates how wave-packet size affects the electron nonlinear resonant interactions. In the case of infinitely long wave-packets ($\beta\to\infty$), resonant electrons with the same initial energies and pitch-angles are divided into two populations. The first population consists of a small group (low probability) of phase-trapped electrons that experience a significant energy gain ($\Delta E>0$). The second population comprises a larger group (high probability) of phase-bunched electrons that undergo a minor energy loss ($\Delta E<0$). For a finite wave-packet size ($\beta=50$), the population of phase-trapped electrons increases due to a higher probability of trapping. For comparison, we also show the $\Delta E$-distribution for the quasi-linear scattering regime (small wave intensity) with scaled $\Delta E\to \Delta E/B_w$. This distribution shows a spread of $\Delta E$ within $[-\max\Delta E,\max\Delta E]$, where $\max\Delta E$ is approximately the energy gain of the trapped population. The significant disparity observed in the $\Delta E$-distributions between the quasi-linear and the nonlinear resonant regime, as well as between the different values of $\beta$ highlights the importance of incorporating realistic wave-packet characteristics into the model of wave-particle resonant interactions around the bow shock.

For the two limits, long wave-packets ($\beta \geq 100$) and very short wave-packets ($\beta \leq20$), there are theoretical approaches modeling the electron distribution evolution in the Earth's radiation belts: systems with $\beta \geq 100$ are described by a generalized kinetic equation (see Refs. \cite{Omura15,Hsieh&Omura17,Hsieh20,Artemyev18:jpp,Vainchtein18:jgr}), whereas systems with $\beta \leq20$ are described by a probabilistic approach (see Ref. \cite{Artemyev21:pre}). The probabilistic approach for $\beta \leq20$ has been adopted for the bow shock wave and plasma characteristics in Paper 1. In this paper, we focus on a generalized approach that can be applied to a wide range of $\beta$. We merge the probabilistic approach and a mapping technique \cite{Artemyev20:pop,Artemyev21:jpp} to describe wave-particle interactions with the wave ensemble having arbitrary distributions in $(B_w,\beta)$ space. 

The paper is structured as follows: Section~\ref{sec:equation} describes the main equations and concepts of nonlinear wave-particle resonant interactions, Section~\ref{sec:long} provides a theoretical model of the mapping technique for very long wave-packets, Section~\ref{sec:short} explains the effects of short wave-packets, Section~\ref{sec:map} generalizes the mapping technique for an arbitrary ensemble of wave-packets, Section~\ref{sec:verification} verifies the generalized mapping technique, and Section~\ref{sec:discussion} discusses the obtained results. 

\section{Hamiltonian equations for resonant systems}\label{sec:equation}
Following the same methodology as in the first paper, we employ two models: one for the bow shock and the other for the foreshock region. Both models describe electron (mass $m_e$, charge $-e$) motion in inhomogeneous magnetic fields with field-aligned whistler mode waves \cite{Artemyev22:jgr:bowshock}:
\begin{eqnarray}
 H &=& \frac{{p_\parallel^2 }}{{2m_e }} + \mu \Omega _0 \left( s \right) - e\Phi \left( s \right) + U_w (s, \mu)\cos \left( {\phi  + \psi } \right) \nonumber\\ 
 U_w &=& \sqrt {\frac{{2\mu \Omega _0 \left( s \right)}}{{m_e c^2 }}} \frac{{eB_w }}{k(s)} \label{eq:01}
 \end{eqnarray}
where $(s,p_\parallel)$ are conjugate variables of parallel coordinate and momentum, $(\psi,\mu)$ are conjugate variables of gyrophase and normalized magnetic moment ($\mu=E\sin^2\alpha/\Omega_0$ where $E$ is the electron energy and $\alpha$ is the electron pitch-angle), $\Omega_0(s)>0$ is the electron gyrofrequency, $\Phi(s)$ is the electrostatic potential describing the polarization electric fields due to ion-electron decoupling around strong magnetic field gradients \cite{Scudder95,Gedalin96}, $B_w$ is the wave amplitude, $\phi$ is the wave phase that determines the wave frequency $\omega=-\partial\phi/\partial t$ and the local wave number $k(s)=\partial\phi/\partial s$. We use the simplified cold plasma dispersion relation $kc=\Omega_{pe}(s)\cdot\left(\Omega_0(s)/\omega-1\right)^{-1/2}$, see Ref. \cite{bookStix62}, where $\Omega_{pe}(s)$ is the plasma frequency. 

The bow shock is described simply by a magnetic field ramp, $\Omega_0(s)=\Omega_{\min}\cdot(1+3b(s))$, a compressional plasma density increase $\Omega_{pe}(s)/\Omega_{pe,\min}\propto\sqrt{\Omega_0(s)/\Omega_{\min}}$, and an electrostatic potential $\Phi(s)=\Phi_0b(s)$ (see, e.g., Ref. \cite{Gedalin96}). Our choice of parameters are typical for the Earth's bow shock: $(\Omega_{pe}/\Omega_0)_{\min}=100$, $\Phi_0\in[0,100]$ eV (see Refs. \cite{Goodrich&Scudder84,Scudder95}). Function $b(s)=(1/2)\cdot\left(1+\tanh(s/L)\right)$ varies from 0 to 1 with a spatial scale $L\approx 1000$ km (the typical scale of magnetic field variations at the bow shock, see Ref. \cite{Krasnoselskikh13}). This spatial scale determines the large system parameter $\eta=L/d_e\sim 10^4$, with $d_e=c/\Omega_{pe,\min}$ being the electron inertial length, the wavelength scale. 

The foreshock magnetic field model corresponds to foreshock transients, localized enhancements of the magnetic field that lead to electron magnetic trapping (see Paper 1 and Refs. \cite{Shi20:foreshock_whistlers,Shi22:ApJ}). This magnetic field model has a magnetic configuration akin to a magnetic field bottle, $\Omega_0(s)=\Omega_{\min}\sqrt{1+(s/L)^2}$, a compressional plasma density increase $\Omega_{pe}(s)=100\Omega_0(s)$, and a large system parameter $\eta=L/d_e\sim 10^4$. We do not include any electrostatic potential mostly because there is no observational study of such potential in foreshock transients. 

\subsection{Hamiltonian of nonlinear pendulum}
The large parameter $\eta$ and small wave amplitude $B_w/B_0\in[10^{-4},10^{-2}]$ (see Refs. \cite{Hull12,Hull20,Shi22:ApJ}) make Equation (\ref{eq:01}) a slow-fast (slow $s$, $p_\parallel$ variables and fast $\phi$, $\psi$ phases) one with a small resonant perturbation $\sim U_w$. To demonstrate the main properties of nonlinear resonant interactions in the Hamiltonian system (\ref{eq:01}), we follow the procedure of typical slow-fast resonant system analysis (see Refs. \cite{Neishtadt06,Neishtadt14:rms,Artemyev18:cnsns}).

First, we introduce phase $\zeta = \phi +\psi$ as a new conjugate coordinate to a new magnetic moment $\tilde\mu$. For this we use a generating function $F(\phi, s; \tilde \mu, p;t) = (\phi +\psi)\tilde \mu + p s$ with new coordinate-momentum pairs $(\tilde s, p)$ and $( \zeta,\tilde \mu )$:
\begin{equation}
    \begin{split}
        \zeta =& \frac{\partial F}{\partial \tilde\mu} = \phi +\psi,\;\; \mu = \frac{\partial F}{\partial \psi} = \tilde \mu,   \\
        \tilde s=& \frac{\partial F}{\partial p} = s, \;\;  p_\parallel = \frac{\partial F}{\partial s} = p + k\tilde \mu. 
    \end{split}
\end{equation}
Because $\tilde \mu = \mu$ and $\tilde s = s$, the tilde sign will be omitted. The new Hamiltonian $\h = H + \partial F / \partial t $ has the form:
\begin{equation}
    \h = \frac{(p+k\mu)^2}{2m_e}+\mu(\Omega_0-\omega)-e\Phi(s)+U_w(s, \mu)\cos\zeta \label{eq:hamiltonian2}
\end{equation}
This Hamiltonian describes a conservative system because $\partial\h/\partial t=0$.

The resonance condition in new Hamiltonian variables is
\begin{equation}
    \dot \zeta = \frac{\partial \h }{\partial \mu} \approx k\frac{(p+k\mu)}{m_e} + \Omega_0 - \omega =0. \label{eq:resonance}
\end{equation}
Equation (\ref{eq:resonance}) determines the resonant momentum $\mu = \mu_R(s, p)$:
\begin{equation}
    \mu_R(s, p ) = \frac{m_e}{k^2}\left(\omega - \Omega_0 - k\frac{p}{m_e}\right).
\end{equation}
Expansion of the Hamiltonian near the resonance gives:
\begin{equation}
    \begin{split}
        \h &\approx \Lambda(s, p)+\frac{k^2}{2m_e}(\mu-\mu_R)^2+U_w(s, \mu_R)\cos\zeta,\\
        \Lambda&(s, p) = \frac{(p+k\mu_R)^2}{2m_e} +\mu_R(\Omega_0-\omega)-e\Phi.
    \end{split}\label{eq:h_approx}
\end{equation}
To introduce the canonical variable $P_\zeta = \mu - \mu_R$, we use the generating function $W(\zeta, s; P_\zeta, \p) = (P_\zeta +\mu_R)\zeta + \p s$ with new conjugate variables:
\begin{equation}
    \begin{split}
        \tilde\zeta =& \frac{\partial W}{\partial P_\zeta} = \zeta,\;\; \mu = \frac{\partial W}{\partial \zeta} = P_\zeta +\mu_R,   \\
        \s=& \frac{\partial W}{\partial \p} = s+ \frac{\partial \mu_R}{\partial \p}\zeta, \;\;  p = \frac{\partial W}{\partial \s} = \p + \frac{\partial \mu_R}{\partial\s}\zeta. 
    \end{split}
\end{equation} 
The first term $\Lambda(s, p)$ in the new Hamiltonian $\tilde \h $ will differ from the first term $\Lambda(s, p)$ from the Hamiltonian (\ref{eq:h_approx}), because of the difference between old variables $(s, p)$ and new variables $(\s, \p)$. Because the second term in the expressions for the old variables is small, it allows us to expand the $\Lambda$ term in the Hamiltonian. Noticing that $s = \s + \{\mu_R,\s  \}\zeta, \, p = \p +\{\mu_R, \p\}\zeta$ where $\{\cdot, \cdot\}$ are the Poisson brackets, it can be shown that
\begin{equation*}
    \Lambda(s, p) = \Lambda(\s,\p) + \{\mu_R, \Lambda\}\zeta,
\end{equation*}
and the new Hamiltonian can be written as:

\begin{equation}
    \tilde H \approx \Lambda(\s, \p) +\frac{k^2}{2m_e}P_\zeta^2+ \{\mu_R, \Lambda\}\zeta + U_w(s, \mu_R)\cos\zeta. 
\end{equation}
In this form, the first term $\Lambda(\s, \p)$ describes the slow motion in the $(s, p)\approx (\tilde{s},\tilde{p})$ plane while the following three are analogous to the nonlinear pendulum Hamiltonian:
\begin{equation}
\h_\zeta   = \frac{1}{{2M}}P_\zeta ^2  + {\rm A}\zeta  + {\rm B}\cos \zeta, \label{eq:hzeta}
\end{equation}
which describes the fast motion near the resonance in the $(\zeta, P_\zeta)$ plane. Coefficients $M$, ${\rm A}$, ${\rm B}$ depend on the coordinates in the $(s, p)$ plane:
\begin{equation*}
    M = \frac{m_e}{k^2} ,\;\;
    {\rm B} = U_w(s, \mu_R) = \sqrt {\frac{{2\mu_R \Omega _0 }}{{m_e c^2 }}} \frac{{eB_w }}{k},
\end{equation*}
and 
\begin{equation*}
    {\rm A} = -\frac{\partial \ln k}{\partial s}\frac{p_{\parallel, R}^2}{k m_e}-\frac{\partial \Omega_0}{\partial s}\frac{p_{\parallel, R}-k\mu_R}{k^2} - \frac ek\frac{\partial \Phi}{\partial s},
\end{equation*}
where $p_{\parallel, R}=m_e\left(\omega-\Omega_{0}\right)/k$ is the solution of equation $\dot\phi+\dot\psi=0$ for the Hamiltonian (\ref{eq:01}) and $\mu_R $ is defined through the electron's resonant energy and the coordinate $s$ of the resonance.

\begin{figure}[!]
\centering
  \begin{subfigure}{0.47\textwidth}
    \includegraphics[width=1\textwidth]{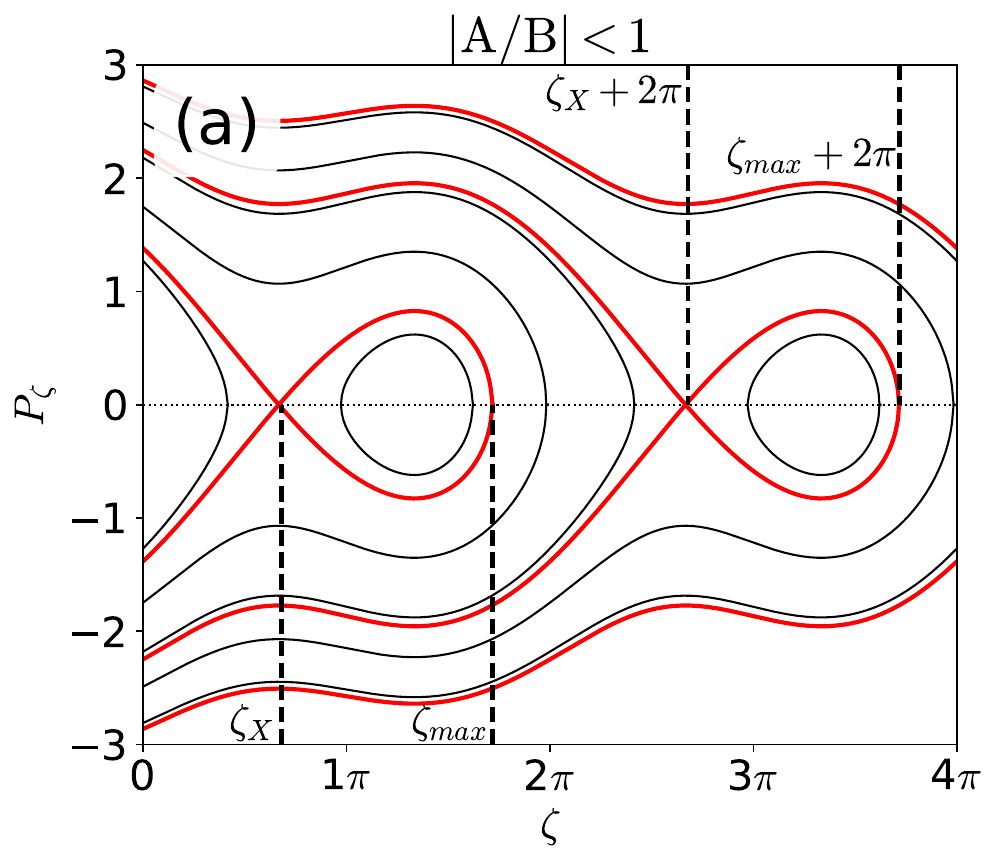}
    \end{subfigure}
  \begin{subfigure}{0.47\textwidth}
    \includegraphics[width=1\textwidth]{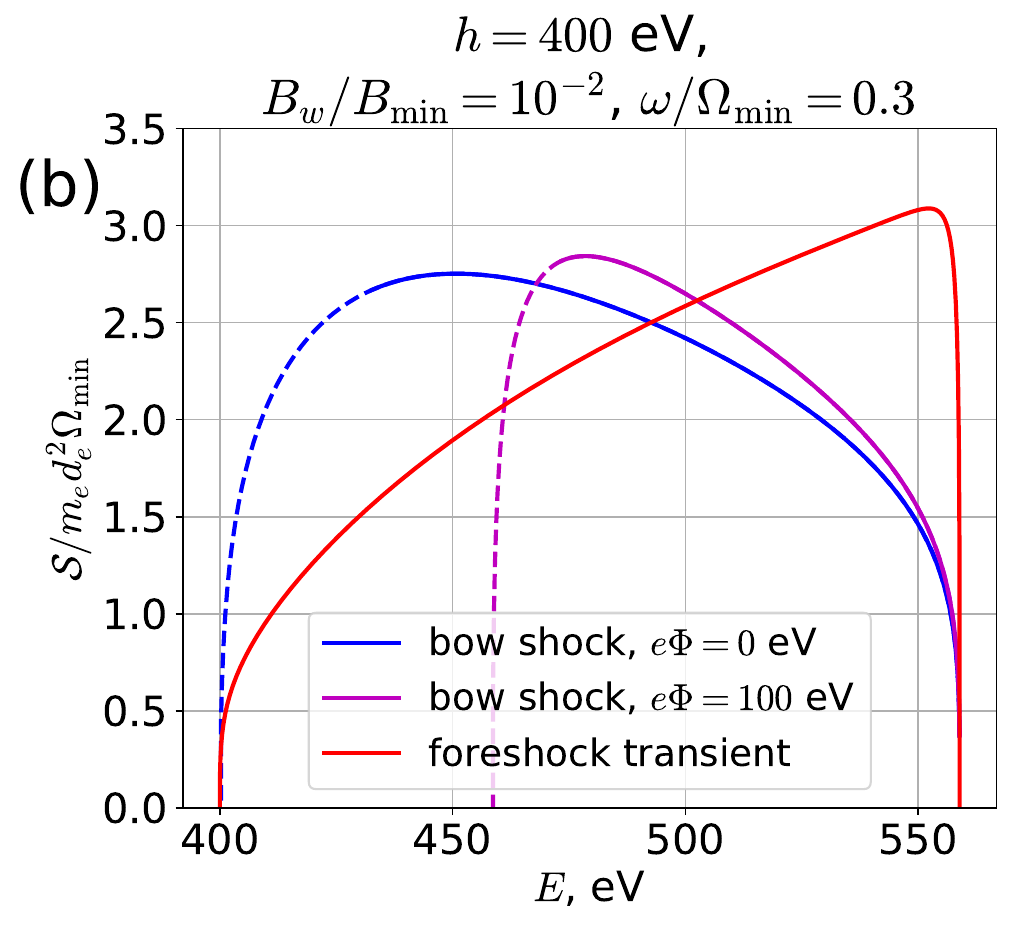}
    \end{subfigure}
\caption{(a) A phase portrait of the Hamiltonian (\ref{eq:hzeta}) with $|A/B|<1$. (b) The separatrix area $\area$ as a function of the initial energy. We plot $\area$ for bow shock and foreshock transient models. For the bow shock model, we trace electrons that move from the upstream region, get reflected from the shock, and then resonate with whistler-mode waves. Dashed lines show the prohibited initial energy range: electrons with such initial energies cannot move away from the shock, because these electrons are not reflected by the shock and go downstream (for a given $h$ value, this corresponds to a certain range of the initial pitch-angle). The maximal resonant energy corresponds to a resonance location ($s$) at the minimum of the ambient magnetic field. 
\label{fig2} } 
\end{figure}

The phase portrait of the Hamiltonian $\h_\zeta$ in $(\zeta, P_\zeta)$ is presented in Fig. \ref{fig2}(top panel). This portrait contains two types of trajectories: {\it transient} trajectories cross the resonance $P_\zeta=0$ once -- electrons moving along such trajectories experience scattering with small energy/pitch-angle change; and {\it phase trapped} trajectories which are closed around the resonance $P_\zeta=0$ -- electrons moving along such trajectories stay around the resonance for a long time (see details in Refs. \cite{Omura91:review,Shklyar09:review,Artemyev18:cnsns}). The trajectory demarcating the phase domains with transient and phase-trapped trajectories is the separatrix. An important system parameter is the area of the region enclosed by the separatrix:
\begin{equation}   
\begin{split}
    \area = 2\sqrt{2M\rm B}\int_{\zeta_X}^{\zeta_{max}}\sqrt{\cos\zeta_X -\cos\zeta+{\rm \frac AB}(\zeta_X - \zeta)}{\rm d}\zeta,  \label{eq:area}
\end{split}
\end{equation}
where $\zeta_X$ is the solution of equations $P_\zeta =0$, $\dot P_\zeta \rightarrow 0$. Area $\area$ depends on $(s,p)$ coordinates in the resonance, where $p=p_{\parallel,R}-k\mu_R$. The combination of the conservation law $E-e\Phi -\omega\mu=h=const$ (see Eq. (\ref{eq:hamiltonian2})) and the resonance condition makes $\area$ a function of energy $E$ only, i.e., $\area=\area(E)$ (see Fig. \ref{fig2}(bottom) showing three examples of $\area(E)$). This function characterizes the electron energy change during resonant interactions, $\Delta E$, and the probability of electron phase trapping, $\Pi$ (see Refs. \cite{Shklyar81,Solovev&Shkliar86,Albert93}). Such probability can be defined as the ratio of resonant electrons that experience phase trapping for a single resonant interaction to the total number of resonant electrons \cite{Neishtadt75,Shklyar81}. 

The initial energy $E$ determines the phase portrait in $(\zeta, P_\zeta)$ plane but cannot describe electron trajectories in the phase portrait, because they also depend on the $\h_\zeta$ magnitude. The coefficients in $\h_\zeta$ depend on $E$, whereas phase $\zeta$ is a fast oscillating variable. Thus, we may introduce an normalized energy $\h_\zeta$ near the resonance $P_\zeta=0$, $2\pi\xi=\zeta_R+({\rm B}/{\rm A})\cos\zeta_R$, and treat $\xi$ as a random variable. An important property of the Hamiltonian (\ref{eq:hzeta}) is that the $\xi$-distribution is uniform (see numerical tests in Refs. \cite{Itin00,Frantsuzov23:jpp}). Therefore, we can use $\area(E)$ and the random variable $\xi$ to characterize the wave-particle resonant interactions \cite{Artemyev18:jpp,Artemyev20:pop}.

\section{Mapping technique for long packets}\label{sec:long}
The $\xi$-averaged characteristics of nonlinear resonant interactions (energy changes due to bunching $\langle\Delta E\rangle_{bunching}$ and trapping $\langle\Delta E\rangle_{trapping}$, and the trapping probability $\Pi$) are determined by the profile $\area(E)$ alone \cite{Artemyev18:jpp,Artemyev20:pop}:
\begin{eqnarray}
 \left\langle {\Delta E } \right\rangle _{bunching}  &=& \omega \left\langle {\Delta \mu} \right\rangle _{bunching} =  - \frac{\omega }{{2\pi}}\area \nonumber\\
 \label{eq:definitions}
 \area\left( {E  + \left\langle {\Delta E } \right\rangle _{trapping} } \right) &=& \area\left( E  \right),\;\;\;  \Pi  =  - \frac{\omega}{{2\pi }}\frac{{d\area}}{{dE}} \nonumber 
 \end{eqnarray}
 with $E-e\Phi-\omega \mu=h=const$. Therefore, knowing $\area$ (see, e.g., Fig. \ref{fig2}(bottom)), we may construct a map for energy changes:
\[
E_{n + 1}  = E_n  + \left\{ {\begin{array}{*{20}c}
   {\left\langle {\Delta E}(E_n) \right\rangle _{bunching} ,} & {\xi  \in \Xi _{bunching}(E_n) }  \\
   {\left\langle {\Delta E}(E_n) \right\rangle _{trapping} ,} & {\xi  \notin \Xi _{bunching}(E_n) }  \\
\end{array}} \right.
\]
where $n$ is the number of resonant interactions (map iteration number), $\Xi_{bunching}(E)$ determines the range of $\xi$ corresponding to bunched particles, and $\xi$ is a random variable that is uniformly distributed over $\xi\in[0,1]$ (see Refs. \cite{Itin00,Frantsuzov23:jpp}). Although the function $\Xi_{bunching}(E)$ can be quite complicated (see details in, e.g., Ref. \cite{Albert22:phase_bunching}), it can be approximated by a simple step-wise function\cite{Artemyev20:pop}:  
\begin{equation}
E_{n + 1}  = E_n  + \left\{ {\begin{array}{*{20}c}
   {\left\langle {\Delta E} \right\rangle _{bunching} ,} & {\xi  \in \left( {\Pi (E_n ),1} \right]}  \\\label{eq:maplong}
   {\left\langle {\Delta E} \right\rangle _{trapping} ,} & {\xi  \in \left[ {0,\Pi (E_n )} \right]}  \\ 
\end{array}} \right.
\end{equation}
This map should be supplemented by the equation of pitch-angle changes ($\alpha_0$ is defined as the pitch angle at the minimum $\Omega_0$): $\alpha_{0,n+1}=\alpha_{0,n}+\Delta\alpha$, with $\Delta \alpha$ determined from $E_{n+1}-e\Phi_R(E_{n+1},\alpha_{0,n+1})-\omega \mu(E_{n+1},\alpha_{0,n+1})=E_{n}-e\Phi_R(E_{n},\alpha_{0,n})-\omega \mu(E_n,\alpha_{0,n})$; here, $\Phi_R(E, \alpha_0) = \Phi(s_R(E, \alpha_0))$ is the electrostatic potential when electrons enter or escape from the resonance (which could be treated as equal for bunching and for trapping and could be found from $\area(E_n) = \area(E_{n+1})$). Iteration number $n$ can be substituted by time as $t_{n+1}=t_n+\tau(E_n)$, with $\tau(E_n)$ being the time interval between two resonance (see Ref. \cite{Artemyev21:jpp}). 

\begin{figure}
\centering
  \begin{subfigure}{0.47\textwidth}
   \includegraphics[width=1\textwidth]{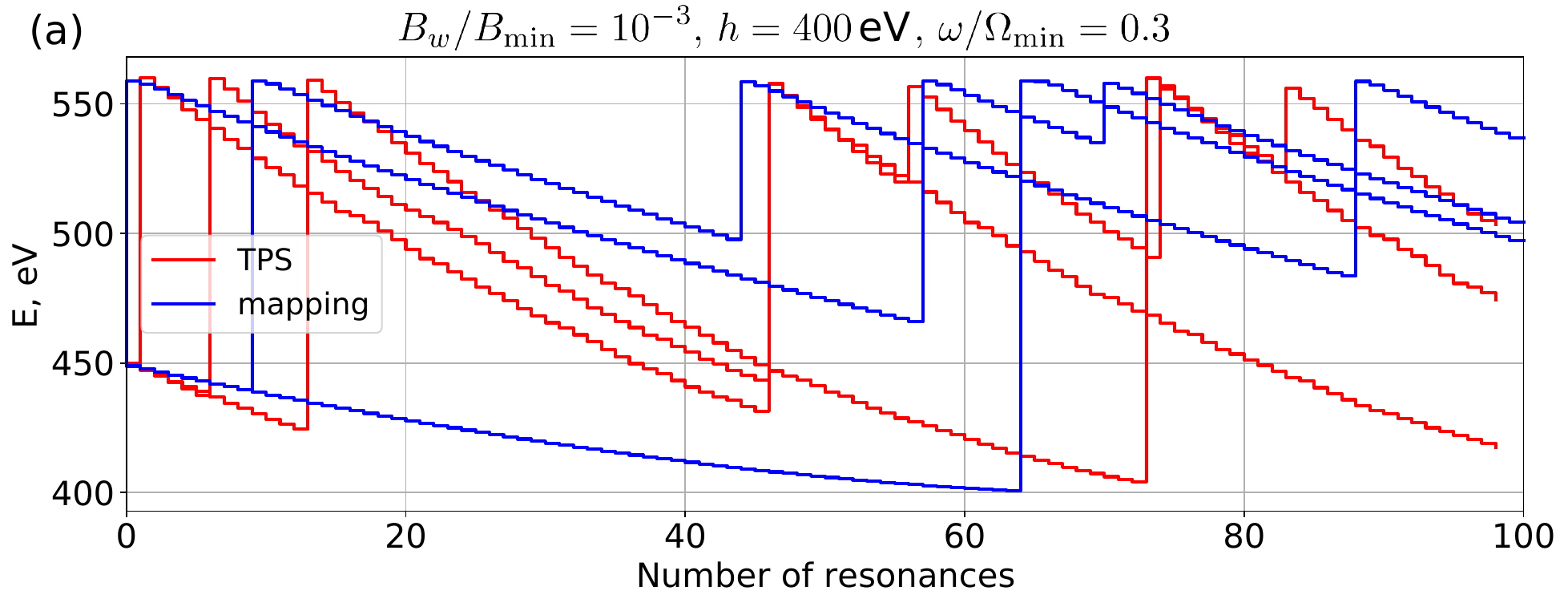}
    \end{subfigure}
  \begin{subfigure}{0.47\textwidth}
   \includegraphics[width=1\textwidth]{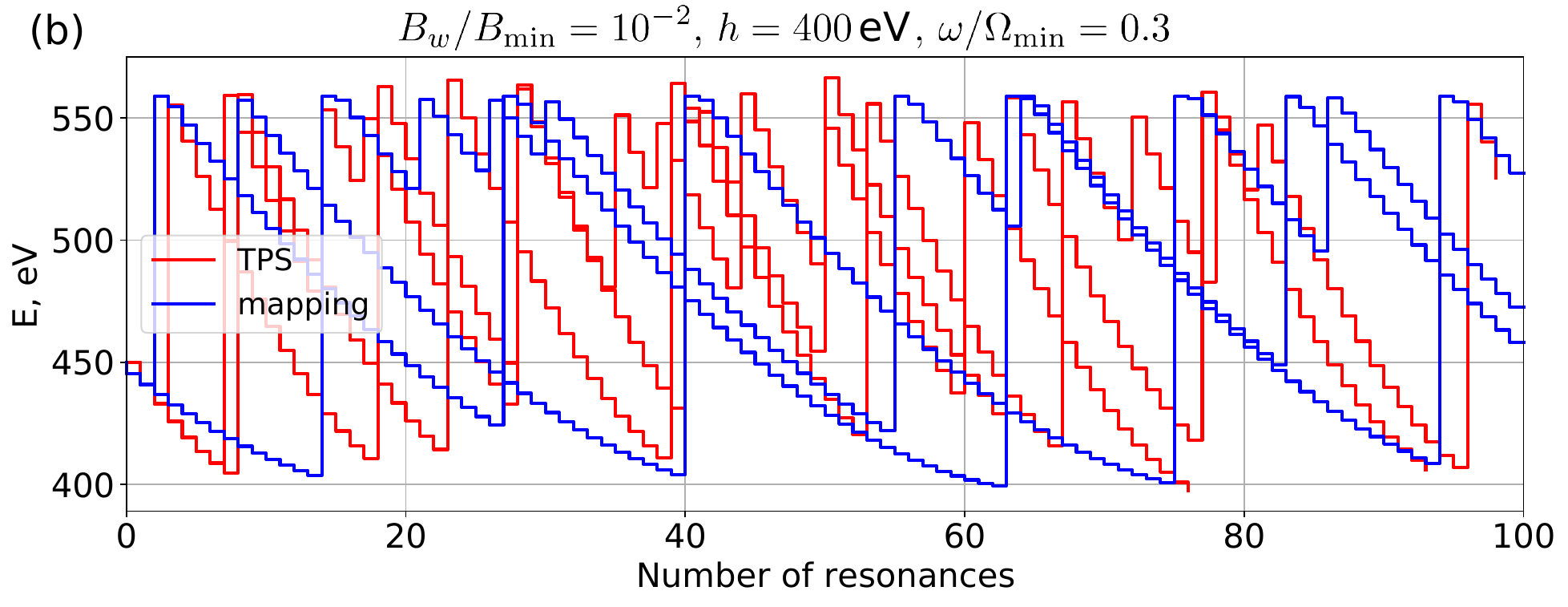}
    \end{subfigure}
\caption{Test particle (TP) trajectories obtained by numerical integration approach for two wave amplitudes (panels a and b, respectively), in comparison with trajectories obtained from the mapping technique (\ref{eq:maplong}). The electron energy is plotted at the resonance as a function of the number of resonant interactions, $n$.
\label{fig3} } 
\end{figure}

Figure \ref{fig3} depicts several examples of electron trajectories calculated by direct integration of the Hamiltonian equation (\ref{eq:01}) and by evaluation of the map (\ref{eq:maplong}). We use the magnetic field model of the foreshock transient to show multiple resonant interactions, and we convert time to the number of resonant interactions $n$. Although the trajectories of electrons with energy $E_n$ are not identical between the results from the test particle approach and the mapping technique (the difference is due to the random $\xi$), these two approaches show statistically similar results: electron energy often decreases due to the phase bunching (small negative jumps) and more rarely increases due to the phase trapping (large jumps). 

\section{Short wave packets}\label{sec:short}
The majority of waves observed around the bow shock are propagating in short wave-packets (or as a train of such short packets), which significantly changes the properties of nonlinear resonant interactions, i.e., the efficiency of phase trapping \cite{An22:Tao,Gan22,Allanson21}. In the first cyclotron resonant interaction, electrons move in the opposite direction to the whistler-mode waves, and thus the packet size controls the time-scale of the electron motion in the trapping regime \cite{Tao13,Zhang18:jgr:intensewaves,Mourenas18:jgr}. However, the probability of trapping depends on $d\area/dE \propto dB_w/ds$ gradient \cite{Neishtadt75,Neishtadt14:rms}, which is very large around the wave-packet edges; this can significantly increase the number of trapped particles \cite{Bortnik08,Artemyev12:pop:nondiffusion,Frantsuzov23:jpp}. Moreover, for all resonance coordinates (i.e., independent of the background magnetic field and plasma gradients), electrons can interact with the wave-packet edge and thus be trapped. Although all these effects seem to complicate the description of wave-particle nonlinear interactions, they also randomize the interaction process and reduce the importance of specific shapes of the $\area(E)$ profile. For sufficiently short wave-packets, $\area(E)$ can be approximated by $\area/2\pi=\omega^{-1}C\cdot \sqrt{\varepsilon}\cdot\left[1-(E/\delta E)^2\right]^{5/4}$, where $\varepsilon=\max B_w/B_0$, $\delta E$ is the resonance energy range for which $\area\ne 0$, and $C\sim 1$ is a numerical factor of the order of one (see derivations of this approximation in Ref. \cite{Artemyev19:pd} and in Appendix of Ref. \cite{Mourenas18:jgr}). Note that we have introduced the normalization factor $\omega^{-1}$ for the $\area$ function, and thus $C$ has a dimension of energy. Interactions with small wave-packets mean that at each interaction we shall use $\area_n/2\pi=\omega^{-1}C\cdot \sqrt{\varepsilon}\cdot\left[1-((E-E_n^*)/\delta E)^2\right]^{5/4}$, with $E_n^*$ determining the location of the wave-packet relative to the electron in resonance. For fixed $h$, the resonant energy $E$ determines the resonant coordinate $s$, and thus the relative location of electrons and wave-packets can be modeled by parameter $E_n^*$ in the $\area_n$ equation. If $E_n \notin [E_n^*-\delta E,E_n^*+\delta E]$, the electron in question will not meet the wave-packet in resonance (see derivations of this approximation in Ref. \cite{Artemyev21:pre}). 

\begin{figure*}
\centering
   \includegraphics[width=1\textwidth]{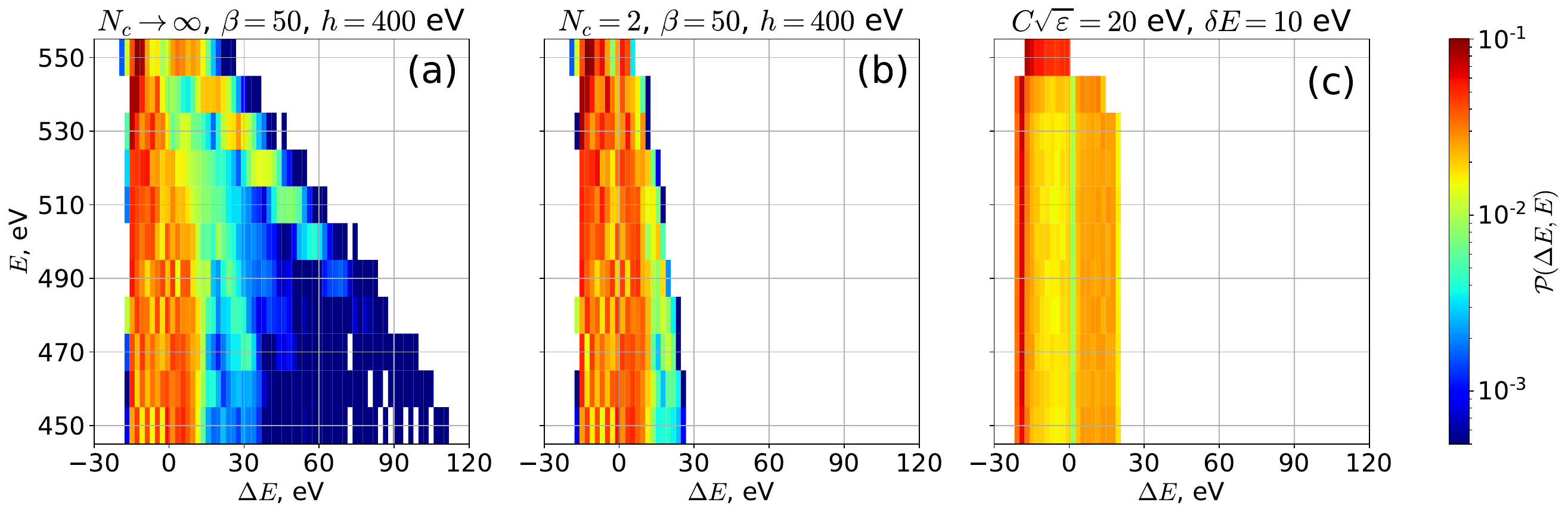}
  
\caption{$\Delta E$-distributions for a single resonant interaction with (a) the coherent wave-packet train, (b) wave-packet train with $N_c=2$. Panel (c) shows $\Delta E$-distribution obtained from the mapping technique. Parameters of map $\area/2\pi=\omega^{-1}C\cdot \sqrt{\varepsilon}\cdot\left[1-((E-E_n^*)/\delta E)^2\right]^{5/4}$ are given on the panel (c), top; see text for details.
\label{fig4} } 
\end{figure*}

To model electron resonant interactions with wave-packets, we numerically integrate the Hamiltonian equations for system (\ref{eq:01}) with $B_w\to B_w f(\phi)$, where $f(\phi)=\exp\left(-5\cos^2(\phi/2\pi\beta) \right)$ is the function modulating the wave field and separating it into wave-packets with the duration of $\sim\beta$ for each packet\cite{Tsai22}. There are two scenarios of electron resonant interactions with the wave-packet train (several wave-packets). The first scenario assumes that the entire train has been generated within the same wave source region that does not change significantly during the wave generation, and thus the phase $\phi$ is coherent along the entire train. In this case, electrons may be retrapped by the next wave-packet after escaping from resonance with the previous wave-packet. This situation is especially common for short wave-packets, i.e., with $\beta<10$, when the distance between two consecutively moving wave-packets is small. Such multi-trapping electron acceleration is quite effective \cite{Hiraga&Omura20,Foster21} and does not principally differ from the electron trapping into an infinitely long wave-packet (except that the trapping probability is higher due to strong wave field gradients at the leading edges of wave-packets). Figure \ref{fig4}(a) shows the distribution of the electron energy change, $\Delta E$, due to resonant interactions with phase-coherent wave-packet trains. The second scenario assumes that wave-modulation (separation of wave field into the wave-packets) destroys the wave phase coherence, and leads to $\phi$ jumps between wave-packets (see examples of in Refs. \cite{Santolik14:rbsp,Zhang20:grl:phase,Nunn21}). Such jumps of $\phi$ prevent the electrons from being trapped into multiple wave-packets \cite{Zhang20:grl:phase}. To model this effect, we introduce a coherence length measured in wave-packet duration, $N_{c}$: all electrons will escape from resonance after interaction with $N_{c}$ packets (some electrons may escape sooner, but no electrons are allowed to stay in resonance longer than this). Figure \ref{fig4}(b) shows the distribution of the electron energy change, $\Delta E$, due to resonant interactions with a wave-packet train having $N_c=2$. Comparison of Panels (a) and (b) demonstrates the effect of wave phase jumps: even for $\beta=50$, the range of $\Delta E>0$ due to trapping significantly shrinks for $N_c=2$, in comparison with $N_c\to \infty$. Note that we use a long wave packet to better illustrate the effect of a finite $N_c$.

We compare these numerically derived $\Delta E$-distributions with the results obtained from electron trajectories integrated using the mapping technique for different $\delta E$ parameters. Figure \ref{fig4}(c) shows that the mapping with $\area_n/2\pi=\omega^{-1}C\cdot \sqrt{\varepsilon}\cdot\left[1-((E-E_n^*)/\delta E)^2\right]^{5/4}$ can reproduce the main properties of the numerically obtained $\Delta E$-distributions: a significant population of electrons do not meet the wave-packet at resonance (peak around $\Delta E\sim 0$). This spreading of the distribution ($\Delta E>0$) is because of the randomization of energy gain of the trapped electrons. Therefore, we propose utilizing a combination of $\area_n/2\pi=\omega^{-1}C\cdot \sqrt{\varepsilon}\cdot\left[1-((E-E_n^*)/\delta E)^2\right]^{5/4}$ functions in the mapping technique to accurately replicate any particular $\Delta E$-distribution obtained from the numerical integration of the electron resonant interactions with wave-packet trains. 

\section{Synthetic map}\label{sec:map}
In this section, we further construct the mapping function as a sum of $\area(E)$ functions. Such a synthetic map allows us to describe the probability distribution $\mP$ in $(\Delta E, E)$ space. We adopt the distribution of wave-packet sizes, $\beta$, as derived from observations (see Fig. \ref{fig1}(a)): $P_\beta=C_0\beta^{-1.67}$ for $\beta\in [2, 100]$ and $C_0=\left(\int P_\beta(\beta)d\beta\right)^{-1}$. Next, we numerically integrate a large ensemble of electron trajectories, where each electron undergoes resonance with wave-packets that have randomly chosen $\beta$ values. 

\begin{figure*}
\centering
  \begin{subfigure}{0.3\textwidth}
   \includegraphics[width=1\textwidth]{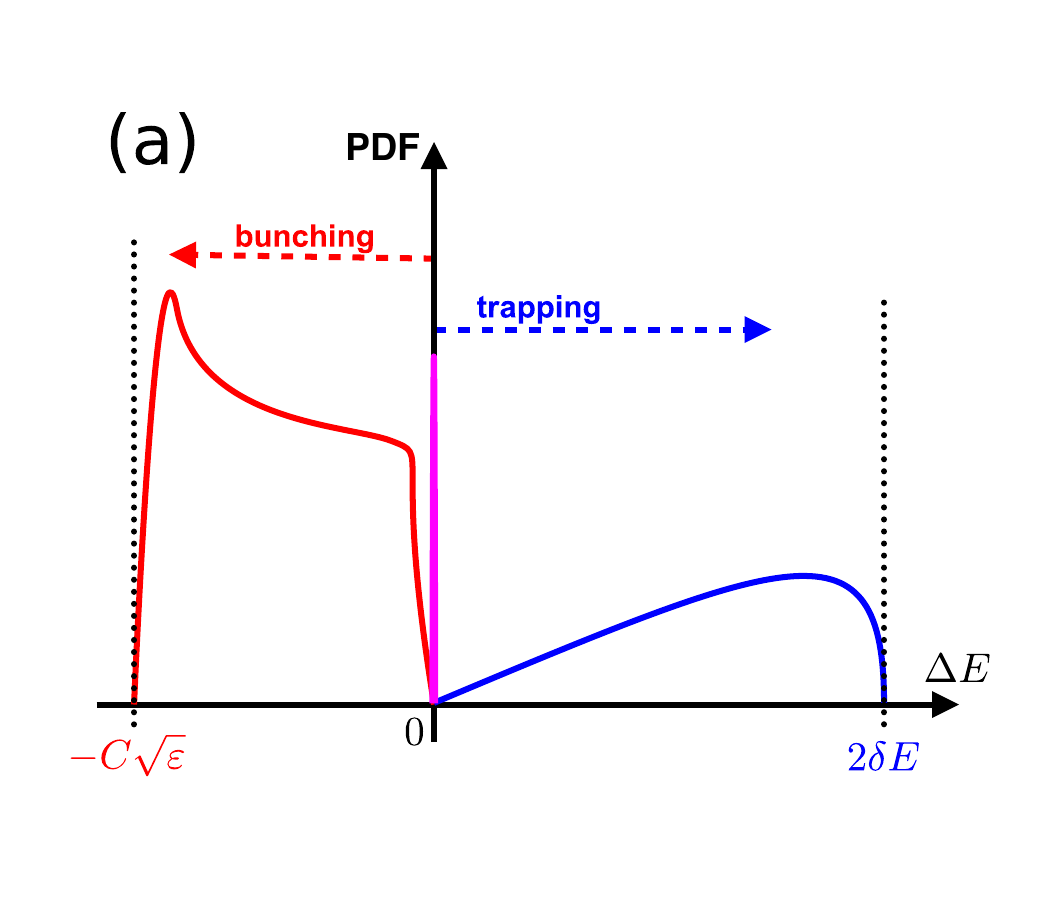}
    \end{subfigure}
  \begin{subfigure}{0.3\textwidth}
   \includegraphics[width=1\textwidth]{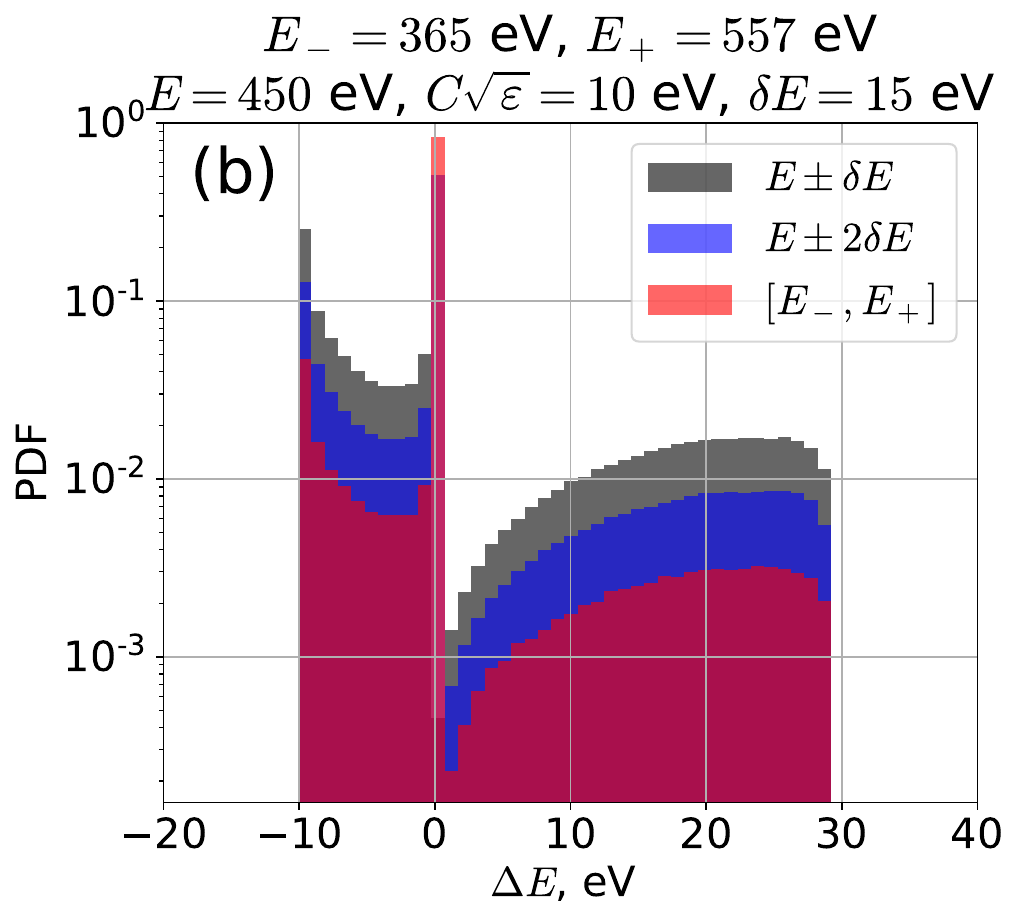}
    \end{subfigure}
    \begin{subfigure}{0.3\textwidth}
   \includegraphics[width=1\textwidth]{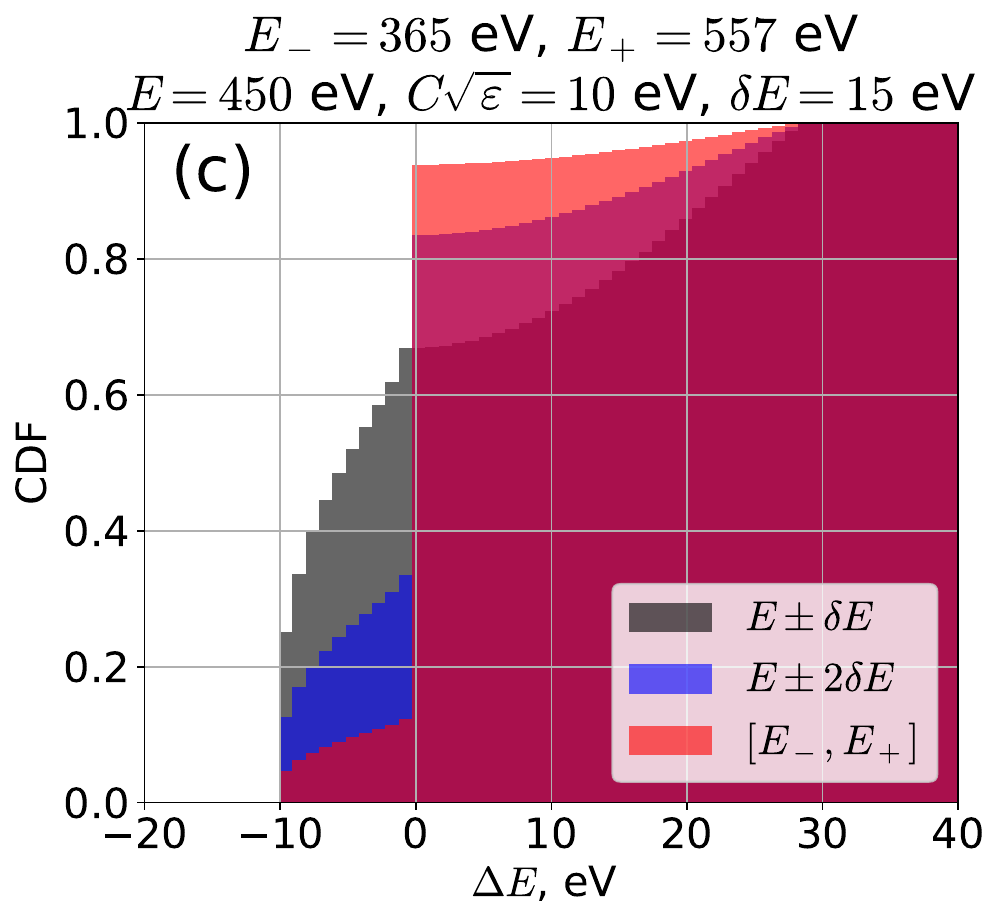}
    \end{subfigure}
\caption{The basic principles to construct the synthetic map: (a) a schematic view of $\delta E$ and $C$ roles in determining the $\Delta E$-distribution characteristics (the resonant energy range is $E\in[E_-,E_+]$ and $\delta E\leq [E_-,E_+]$); (b) three $\Delta E$-distributions for three uniform $E_n^*$ distributions within $E\pm\delta E$ (black), within $[E_-,E_+]$ (red), and within $E\pm2\delta E$ (blue); (c) cumulative distribution functions for $\Delta E$-distributions from (b). \label{fig5} } 
\end{figure*}

To construct such a synthetic map (a sum of many maps with different $\area(E)$), we first determine how $\area(E)$ parameters (i.e., $\delta E$, $C$, $E_n^*$ distributions) control the resulting $\Delta E$-distribution. Figure \ref{fig5}(a) depicts a schematic of the role of $\delta E$ and $C$ in determining the $\Delta E$-distribution characteristics. The magnitude of $\area(E)$ controls the $\Delta E$ range of bunching with $\min \Delta E=-\max\omega\area(E)/2\pi$, whereas $\delta E$ controls the $\Delta E$ range of trapping with $\max \Delta E=\delta E$. The maximum relative number of trapped particles, $\Pi$, is determined by $\max \Pi/\sqrt{\varepsilon} = (C/\delta E)\cdot (5/2^{1/2}3^{3/4})\approx 1.551\cdot(C/\delta E)$. Both bunching and trapping energy ranges $\Delta E$ depend on initial electron energy $E$ (see Fig. \ref{fig4}). As a result, we set $C=C(E)$ and $\delta E=\delta E(E)$. If such dependencies are weak, $|dC/dE|\ll C/\delta E$ and $d\delta E/dE \ll 1$, they will not significantly change the value of $\Pi$. 

The distribution of wave-packet positions, $E_n^*$, determines how often electrons will resonate with waves. If $E_n^*$ is uniformly distributed within $E\pm\delta E$, then each map iteration will have an energy change, i.e., electrons will encounter wave-packets each time when crossing the resonance. This describes a system with dense wave-packet trains, where one packet moves right after another. If $E_n^*$ is uniformly distributed within the entire resonant energy range $[E_-,E_+]$ and $\delta E<E_+-E_-$, then the probability of the particle energy change (the probability to meet a wave-packet in resonance) is $\delta E/(E_+-E_-)$. For small $\delta E/(E_+-E_-)$, the resonant interactions are rare, and this describes a system with well-separated individual wave-packets. Figure \ref{fig5}(b) shows $\Delta E$-distributions for three different $E_n^*$ distributions: a smaller probability of resonant interactions means a higher value of probability to have $\Delta E=0$ (see a peak of $\mP$ at $\Delta E=0$).

\subsection{Single resonance system}
\begin{figure}
\centering
\includegraphics[width=.5\textwidth]{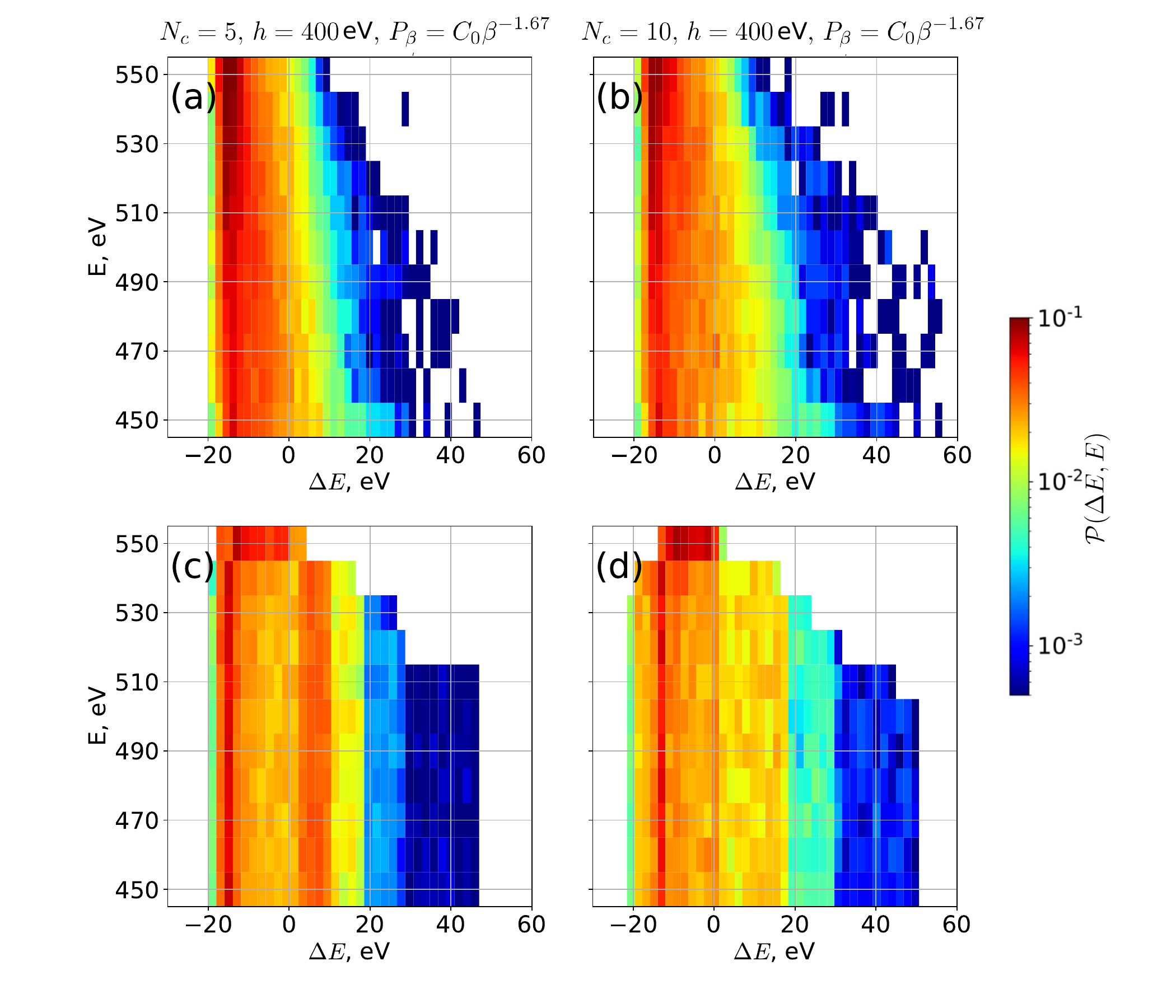} 
\caption{Panels (a,b) show two distributions $\mP (\Delta E, E)$ for the bow shock magnetic field model and $N_c=5$(a), $N_c=10$(b). For each resonant interaction, we use a random value of $\beta$ determined by the probability distribution function $P_\beta=C_0\cdot \beta^{-1.67}$ for $\beta\in[2, 100]$ and $C_0=\left(\int P(\beta)d\beta\right)^{-1}$ and the same wave amplitude $B_w/\min B_0=10^{-2}$. Panels (c,d) show two distributions $G(\Delta E, E)$ obtained from the mapping of a sum of $\area_k$ with $P_{\delta E }(\delta E_k) = a_k \cdot (\delta E)^{-1.67}$. For panel (c) $\delta E_k =\{1,9, 15, 25\}$, $\sqrt{\varepsilon}C_k =  \{9,14,18, 22\}$,  $a_k = \{6.3, 17.7, 5.7, 9.4\}$, and for panel (d) $\delta E_k =\{5,9, 14, 23\}$, $\sqrt{\varepsilon}C_k = \{17,15,22, 19\}$,  $a_k = \{0.1,22.0,22.0,22.0\}$.
\label{fig6} } 
\end{figure}

Figure \ref{fig6}(a,b) shows the probability distribution functions $\mP (\Delta E, E)$  for a single resonant interaction with randomly chosen $\beta$, $N_c=5$ and $N_c=10$. For $N_c=10$, the distribution $\mP (\Delta E, E)$ has a slightly higher probability for large $\Delta E>0$ at low energies ($E\leq 490$eV), but the difference between $N_c=5$ and $N_c=10$ is significantly reduced due to a wide $\beta$ distribution dominated by small $\beta$ (see Fig. \ref{fig1}). For $N_c=5$, the distribution $\mP (\Delta E, E)$ is almost independent of $E$. In the first paper, we have examined the limit of small $N_c$ when the probability distribution function $\mP (\Delta E, E)$ can be reduced to 1D $\mP (\Delta E)$, without significant dependence on $E$.

We fit two $\mP(\Delta E, E)$ distributions from Fig. \ref{fig6}(a,b) by $G(\Delta E, E)$ functions derived from the synthetic map consisting of a sum ($\sum_k \area_k P_{\delta E}(\delta E_k)$) of 
\[\area_k/2\pi=\omega^{-1}C_k\cdot \sqrt{\varepsilon}\cdot\left(1-\left((E-E_n^*)/\delta E_k\right)^2\right)^{5/4}\]
We use $P_{\delta E}\propto(\delta E)^{-1.67}$, which is based on the $P_\beta$ distribution and the $\delta E\propto\beta$ relation (see schematic in Fig. \ref{fig5}  and Ref. \cite{Artemyev21:pre}). Figure \ref{fig6}(c,d) shows two $G(\Delta E, E)$ distributions constructed with four $\area_k$ functions having different $C_k$ and $E_n^*=E\pm\delta E_k$ (parameters are in the figure caption). These synthetic $G(\Delta E, E)$ distributions reproduce the main details of the numerically obtained $\mP (\Delta E, E)$ distributions. Importantly, the fitting procedure for $G(\Delta E, E)$ does not have a unique solution, but should approach the distribution of energy changes $\Delta E$ that fully describes electron resonant dynamics. Thus, we will further test whether the constructed  $G(\Delta E, E)$ provides a correct long-term evolution of the electron ensemble (see Sect.~\ref{sec:verification}).

\subsection{Magnetically trapped electrons}
\begin{figure*}
\centering
   \begin{subfigure}{0.4\textwidth}
   \includegraphics[width=\textwidth]{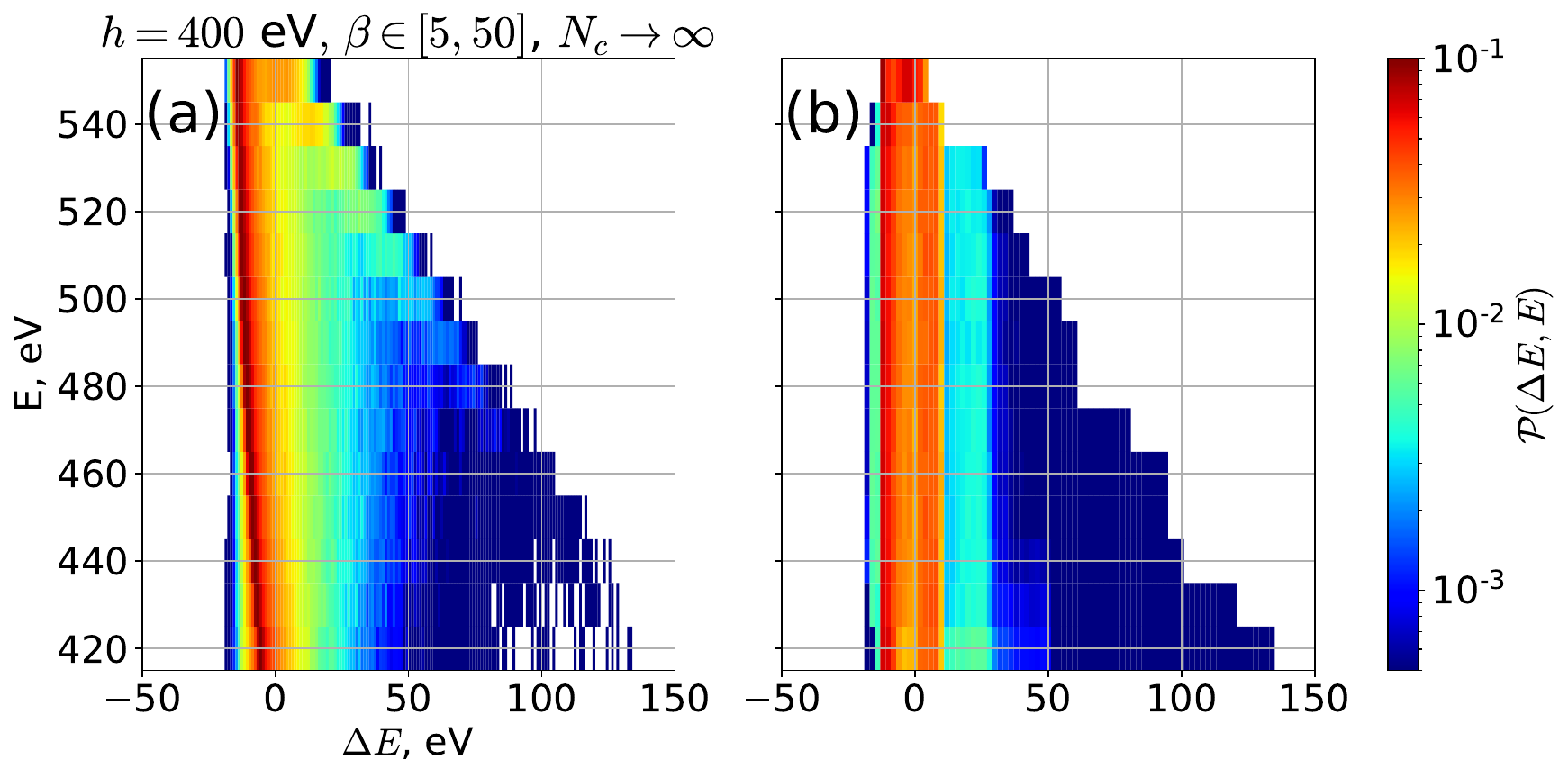}
    \end{subfigure}
    \begin{subfigure}{0.4\textwidth}
   \includegraphics[width=\textwidth]{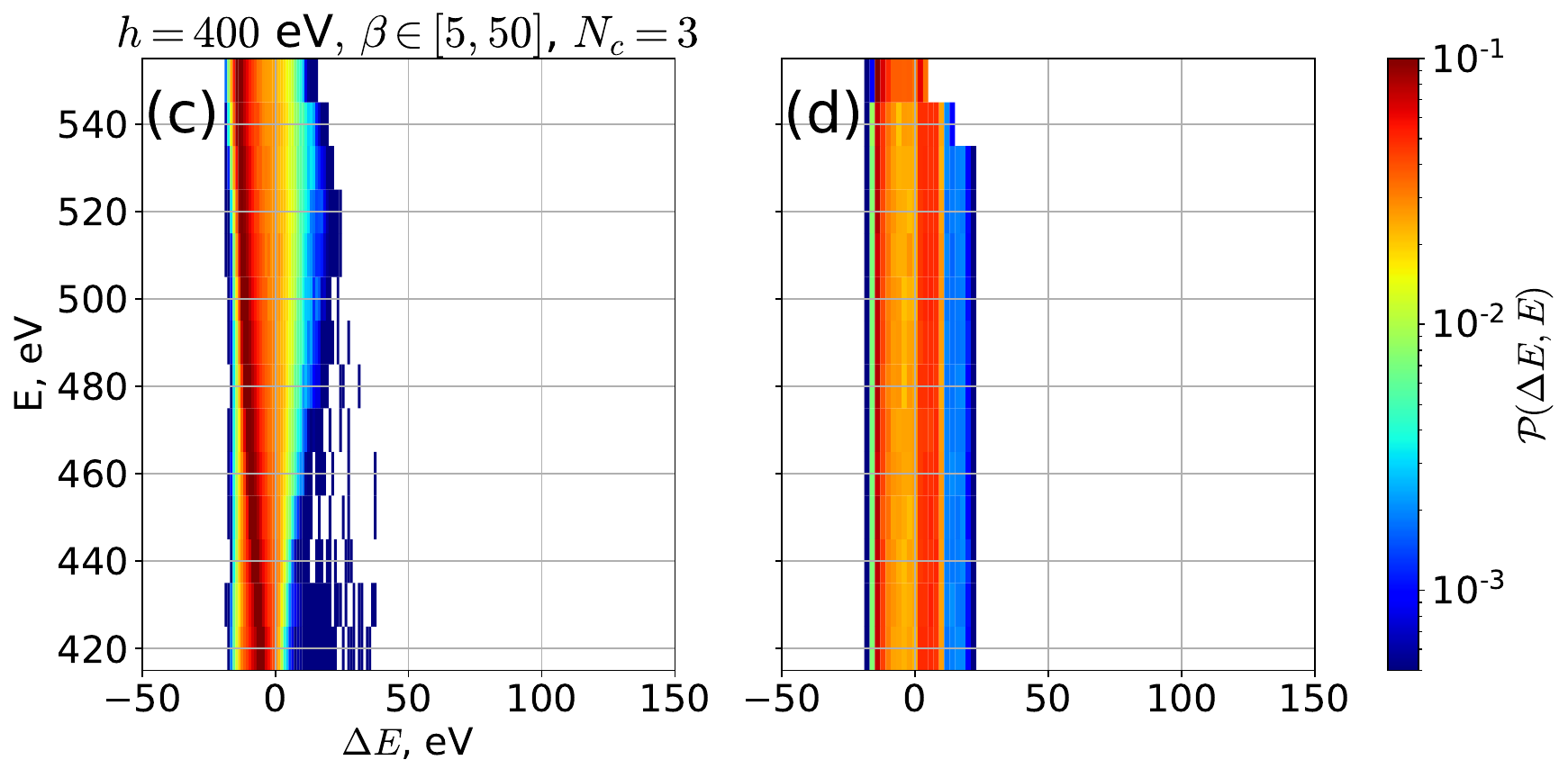}
    \end{subfigure}
    
    \begin{subfigure}{0.4\textwidth}
   \includegraphics[width=\textwidth]{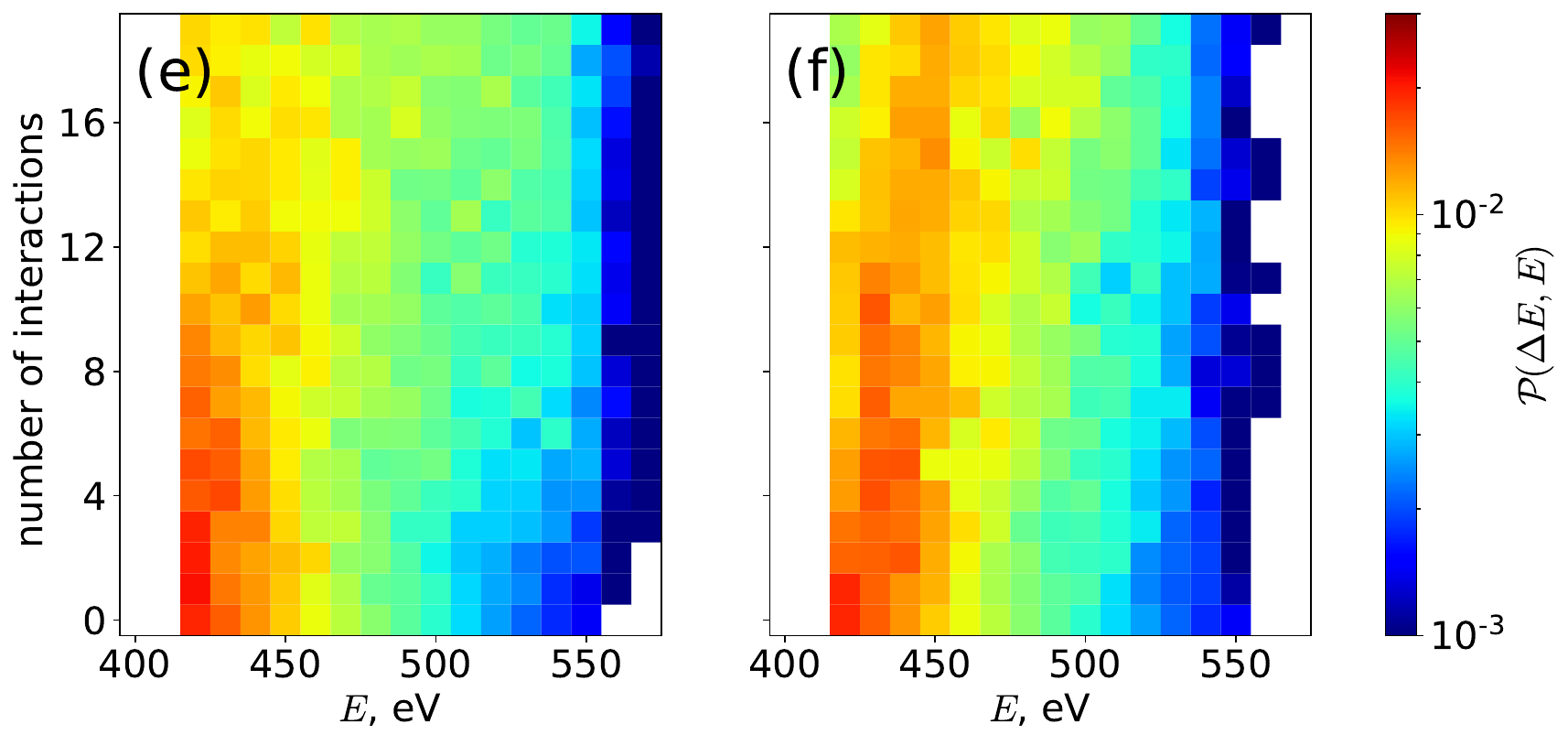}
    \end{subfigure}
\centering
  \begin{subfigure}{0.4\textwidth}
   \includegraphics[width=\textwidth]{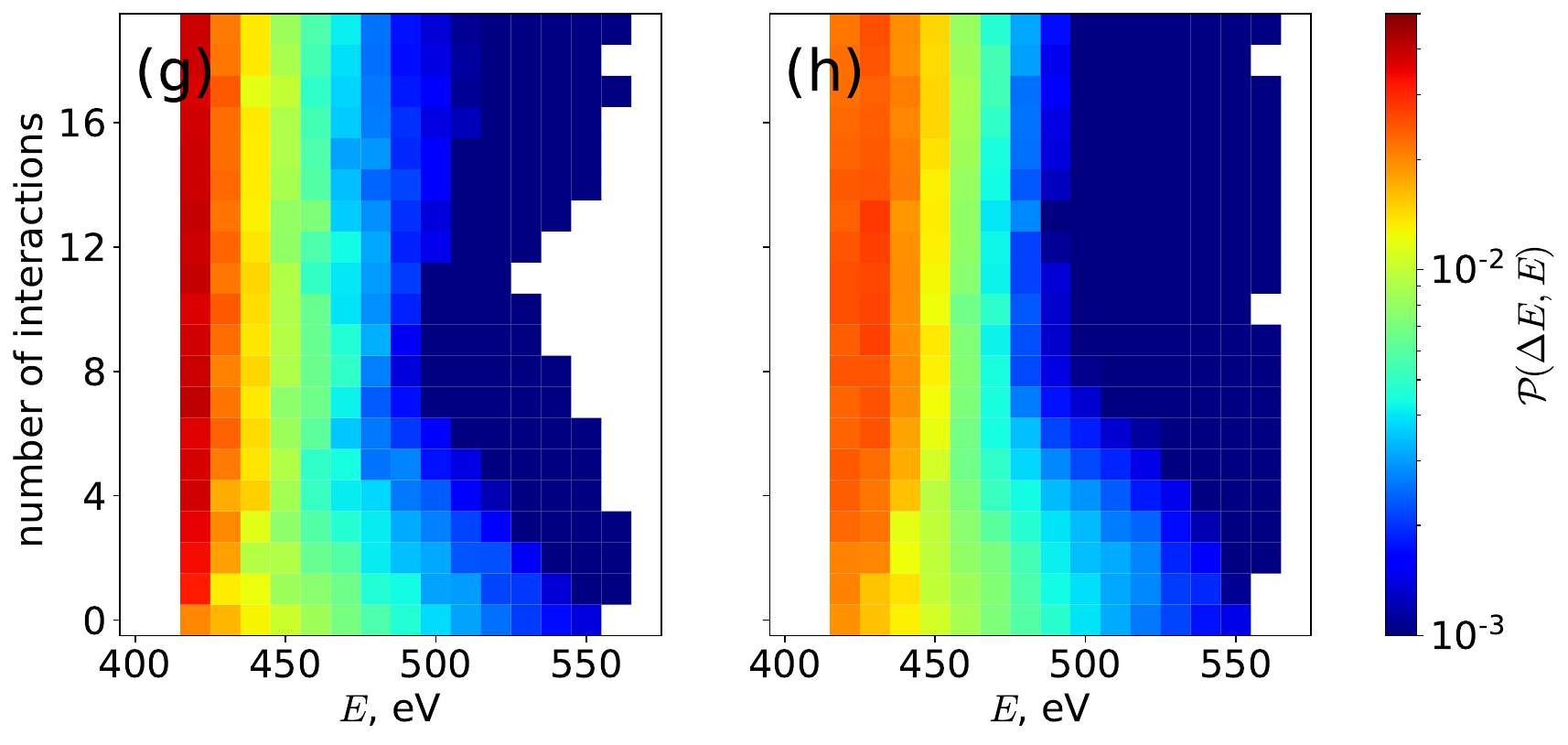}
    \end{subfigure}
 
\caption{Panels (a, b) show the probability distribution function $\mP (\Delta E, E)$ for $N_c \rightarrow \infty $ and the corresponding synthetic map, starting from the initial distribution of $\sim\exp\left(- E/50 \text{ eV}\right)$; panels (c, d) show the same but for $N_c =3$. Parameters for (b): $\delta E_k =\{67, 47,30, 26, 18, 15, 14, 5\}$, $\sqrt{\varepsilon}C_k =  \{10, 16, 20, 16, 14, 8, 11, 12\}$,  $a_k = \{3.7,  7.8,  2.9,  7.7,  2.5, 6.1, 10.3, 10.7\} $; for (d): $\delta E_k =\{11, 10, 5, 4\}$, $\sqrt{\varepsilon}C_k =  \{21, 17, 14, 12\}$,  $a_k = \{0.4, 3.1, 4.9, 6.0\}$. Panels (e, f) show the evolution of the energy distribution with $N_c \rightarrow \infty$ and the corresponding evolution of distribution with the use of the synthetic map from panel (b); panels (g, h) show the same but for $N_c = 3$ with the synthetic map from panel (d).
\label{fig7} } 
\end{figure*}

Figure \ref{fig7}(a,c) repeats the results from Fig. \ref{fig6}(a,b) for a single resonant interaction with wave-packets having $\beta$ distributions of $P_\beta\propto\beta^{-1.67}$. However, in this case, the trajectories are traced in the foreshock transient magnetic field model. For $N_c\to\infty$, the shortness of wave-packets can be compensated by the effect of multiple trapping, and the corresponding probability distribution function $\mP (\Delta E, E)$ contains a finite probability of large $\Delta E>0$ for small $E$. Despite that the $\Delta E$-distribution depends on $E$, the synthetic mapping technique reproduces the main features of the 2D distribution $\mP (\Delta E, E)$ (compare Fig. \ref{fig7}(a) and Fig. \ref{fig7}(b)). 

For $N_c=3$, there is almost no effect of multiple trappings, and the corresponding probability distribution function $\mP (\Delta E, E)$ does not contain large $\Delta E>0$ jumps. Moreover, due to the dominant role of small $\beta$ in $P_\beta\propto\beta^{-1.67}$ distribution, the energy change $\Delta E$ due to resonant interactions only weakly depends on the initial energy $E$ (similar to the bow shock system; compare Fig. \ref{fig7}(c) and Fig. \ref{fig6}(a)). Although the synthetic map can describe such a 1D distribution of $\mP (\Delta E)$ (compare Fig. \ref{fig7}(c) and Fig. \ref{fig7}(d)), we note that a simpler approach for such $E$-independent systems has been proposed in Paper 1.

\section{Verification of the mapping technique}\label{sec:verification}
We next verify the suggested approach for constructing synthetic maps in the foreshock transient magnetic field model. This assumes multiple resonant interactions for each electron, and thus includes the long-term dynamics of the electron energy distribution. In the test particle simulation, we numerically integrate $10^4$ electron trajectories with the same $h$ and initial energy distributions with energy falloff $\sim \exp(-E/50{\rm eV})$. Trajectories are trapped in the system with an ensemble of wave-packets described by $P_\beta$, but for half of the bounce period each electron only resonates with one wave-packet (new wave characteristics for the next half of the bounce period are set when an electron crosses the $B_0$ minimum). Electrons that undergo phase bunching tend to drift towards smaller energy and pitch angles. This drift may cause them to enter the small pitch-angle range, where the mapping technique should be modified due to the effects of anomalous trapping \cite{Kitahara&Katoh19}. Moreover, such electrons should escape the system because their mirror points are outside of the foreshock transient region (magnetic field magnitude in the mirror points exceed $\sim 10\times$ of the minimum magnetic field). Thus, we exclude electrons reaching the small pitch-angle boundary, $\alpha_0\approx 20^\circ$. To keep the same number of particles during the simulation, we substitute each escaping electron with a new one (e.g., trapped into the foreshock transient from the solar wind flow) of identical initial energy as the lost electron. Modeling them this way is equivalent to including a large trapping probability for small energy electrons, to reflect them from the small-energy/small pitch-angle boundary (see Refs. \cite{Kitahara&Katoh19,Albert21,Artemyev21:pop} for discussions of this effect for electron scattering by whistler-mode waves in the radiation belts).

Figures \ref{fig7}(e,f) and (g,h) compare the results of test particle simulations and the mapping technique using the synthetic maps shown in Fig. \ref{fig7}(b) and (d), respectively. For a large $N_c$, we see a significant acceleration of electrons, whereas, for small $N_c$, electrons mostly drift to small energy/pitch-angle ranges and form a field-aligned population. These effects are clearly evident in the results obtained from the test particle simulations (panels (e, g)) as well as in the mapping technique results (panels (f, h)). Thus, the presented comparison validates the proposed mapping technique and synthetic map. 

\section{Discussion and Conclusions} \label{sec:discussion}

This is the second paper of two accompanying studies on modeling the nonlinear wave-particle resonant interactions around the Earth’s bow shock. Paper 1 described the probabilistic approach for resonant interactions dominated by short wave-packets, while this (second) paper merges the probabilistic approach and mapping technique to describe the rare but effective electron acceleration by long wave-packets. This generalized approach is based on the generation of a synthetic map, which reproduces the main statistical properties of resonant wave-particle interactions, while keeping all important relations analytical during the modeling of such interactions. Below is a summary of our results::
\begin{itemize}
  \item We have demonstrated the technique of constructing a synthetic map that superposes multiple analytical maps describing nonlinear resonant interactions.
  \item The main parameters of such a synthetic map can be determined by comparing mapping results and the probability distribution function of electron energy changes, derived, e.g., from test particle simulations.
  \item The synthetic map approach allows the modeling of the long-term dynamics of electron distributions with multiple nonlinear resonant interactions with realistic wave-packet distributions in amplitude and size. 
\item The proposed synthetic map has been verified with test particle simulations for the magnetic field model in Earth's foreshock transients.
\end{itemize}

This approach can be useful for modeling the electron resonant interaction effects in large-scale simulations, where the long-term electron dynamics is dominated by adiabatic heating and transport, whereas wave-particle resonant effects should be responsible for scattering and relaxation of the electron anisotropy (see discussions in Refs. \cite{Balikhin93,Schwartz11:VladimirShock,See13,Wilson14,Mozer&Sundkvist13,Gedalin20}). Although we have developed and verified this approach for electron resonant interactions with whistler-mode waves, the same effects of nonlinear resonant phase trapping and bunching can be found in systems of intense electrostatic waves \cite{Artemyev19:cnsns} and ion/electron holes \cite{Vasko18:grl}. Moreover, this approach can be included in test particle simulations of electron heating by low-frequency compressional waves in the presence of whistler-mode wave scattering (i.e., the so-called pumping effect, see Refs. \cite{Lichko17,Borovsky17:pumping,Fowler20}). Systems with such a combination of adiabatic heating and wave-particle scattering, which controls the long-term dynamics of electrons, are not limited to just the solar wind or shocks: a similar interplay of large-scale adiabatic processes and small-scale wave-particle interactions are believed to control electron fluxes in planetary radiation belts \cite{Elkington18:agu,Elkington19:agu} and reconnection outflow regions \cite{Ukhorskiy22,Artemyev22:jgr:DF&ELFIN}. Therefore, the proposed synthetic map can be further combined with test particle simulations in these systems. 

\section*{Acknowledgments}
X.S., A.V.A., X.-J.Z., and V.A. acknowledge THEMIS contract NAS5-02099, NASA grants 80NSSC22K1634, 80NSSC21K0581 (spacecraft data analysis and numerical simulations). A.V.A. and D.S.T. also acknowledge Russian Science Foundation through grant No. 19-12-00313 (theoretical models).

\section*{Data Availability}
This is a theoretical study, and all figures are plotted using numerical solutions of equations provided in the paper. The data used for figures and findings in this study are available from the corresponding author upon a reasonable request. Spacecraft data for the first figure are publicly available in MMS data repository \url{https://lasp.colorado.edu/mms/sdc/public}.  Data access and processing were done using SPEDAS V4.1 \cite{Angelopoulos19}.

\bibliographystyle{unsrtnat}

\end{document}